# From Self-Assembly to Controlled-Assembly
# From Optical Manipulation to AFM Manipulation

Farbod Shafiei*

Department of Physics

The University of Texas at Austin, Austin TX 78712

*farbod@physics.utexas.edu

Moving nanoparticles/atoms to study the nearfield interaction between them is one of the many approaches to explore the optical and electrical properties of these assemblies. Traditional approach included the self assembly by spinning or drying nanoparticles in aqua on the substrate is well practiced. Lithography technique is another popular approach to deposit limited nano/micro patterns on substrates. Later optical and mechanical manipulations were used to have more control over moving individual elements of nano and microstructures and even atoms. Optical tweezers, optical trapping and AFM manipulation are examples of these precise approaches.

## Introduction

Proposing and building optical nano-circuits based on manipulating local optical electricfield by using photon rather than electron [1, 2], bring us to the point that precise building of nanostructure is playing big role in our future nano opto-electronic science.

Recently the older concept of optical standing wave for creating array of micrometer size polystyrene spheres [3] was used for optical trapping a mirror structure made of micro-beads. T. Grzegorczyk et al. [4] used laser light for trapping array of micrometer size polystyrene beads acting as a mirror. Although this laser trapped mirror was in water, against chamber wall and was not fully controlled to create curved mirror but it's a very strong step toward practical implementation of optical device made of micrometer size particles in space [5].

Using scanning probe microscopes for mechanical manipulation of atoms, molecules, micro or nanoparticles has been established previously. Atomic force microscope (AFM), nearfield scanning optical microscope (NSOM) or scanning transmission microscope (STM) were used in these process by attaching the particles to the probe and move the particles around. In different approach, atom, molecule, micro and nanoparticles were moved around and manipulated by the probes of these systems on substrate.

D. Eigler et al. [6] in their famous work were able to move xenon atoms in a controlled way by STM probe and build the IBM logo out of individual atoms on nickel surface. By keeping the right distance and voltage between the STM probe and xenon atom, they were able to drag single atom across the substrate.

Using the scanning probe systems has this advantage of being able to move the particles that are attached to the probe while observing the interaction of them with other particles and elements of sample [7, 8] or manipulating and enhancing the interaction between elements [9, 10] .

The assembly of nanoparticles became so important that they were claimed as future atoms and molecules of tomorrow's materials if they can be successfully assembled into useful structures [11]. The properties and possible applications of these assemblies rely on ability of controlling the interactions between the electronic, magnetic and optical properties of the individual nanoparticles [12]. This brings us to the point that the precise assembly of atoms, molecules and nanoparticles are essential to obtain and observe desired properties.   Here we look very briefly at and beyond some of the self assembly techniques that do not have enough precision for this purpose. We briefly looked at other precise manipulation technique as well.

## A. Self Assembly

Self-assembly of nanoparticles refers to the process that these building blocks spontaneously self-organize themselves through the process. As an example they could land side by side in close pack configuration or get attached through DNA or ligands to each other. This is considered as bottom up approach.

Self assembly of quantum dots to superlattices is done by close packing them via gentle evaporation of liquid in the form of ordered or oriented thin film or structure [13, 14]. This type of closed packed array of nanoparticles could be linked covalently by molecules as well [15].

Since early 90's, variety of shapes became available for nanoparticles [11, 16] and variety of structure have been reported from self assembly of nanoparticles, including chain, sheet, vesicles, 1, 2 and 3 dimensional crystals and …. [12, 17, 18].

## B. Controlled Assembly

### 1. DNA and Template Assembly

Later those self assembly of nanoparticles were targeted to specific arrangement like DNA assembly of nanoparticles, from simple aggregation to ordered aggregation leading to network of nanoparticles [19] or discrete assembly [20]. Specific arrangement of nanoparticles connected to each other by DNA were achieved later in dimer or trimer configuration [21] or even pyramid configuration [22].

By having a template and active area of the substrate, there was more control to arrange the elements on specific area of the substrate due to chemical active region. Later elements such as nano and microparticles were assembled on this active area. Template pattern has been reported from simple discrete array [23] up to triangle origami [24] and even carbon nanotube template [25].

In template assembly, still it will be hard to have fine control over the desired configuration and specially the distance between the particles and elements. B Yan et al. [26] work is an example of poor control over assigning same number of nanopartilces to the active template and almost no control over the distance between them. Micro-molecular template self assembly gives more control over the assembly [27].

External parameter such as electric or magnetic field, light, flow, temperature, PH…were used to do more fine tuning in these type of template assembly [28].

### 2. Photolithography Assembly

Other approach to have nano and micro structure in desired configuration is photolithography and deposition assembly or similar approaches. In this approach photoresist and patterned mask is used to create a pattern on substrate to later be filled by deposition technique and clear off the unwanted pattern by lift off. The advantage of this approach is that there are more control, up to system resolution, of how close the nano and microstructures would be at the time of deposition. The components could have almost any shape and configuration but due to deposition technique, it is not possible to have curved surface and edges. Imagine replacing sphere by a cylinder when you switch from nanoparticles self assembly technique to photolithography technique.

Plasmonic optical study of of oligomers pattern [29] by photolithography shows how the configuration and distance between the elements are under the control in this approach. This method has been used for overlapping structures as well [30].

In addition to the restriction of type of material and the shape of structure due to nature of deposition in this technique, the built structure could not be modified. This problem also exists in self assembly with and without template as well. To go around such an issue you have to make several structures to replicate the intermediate configurations, if it is need to be study. In M Hentschel et al. [29] work, it is clear that variety oligomers has been built with different gap between nanoelements to show the difference between optical studies of these structure. Although these nano elements and structures are very similar but you end up comparing different structures to each other as you can not modify the original structure. If such a thing is crucial for the experiment, the more elaborate technique such as optical or mechanical manipulation is needed.

## 3. Optical Manipulation

### 3-1. Optical Tweezers

Optical manipulation technique that was originally proposed for atom trap [31] is based on single beam optical trap known as optical tweezers [32]. A highly focused beam and the presence of intensity gradient force can drag small particles in aqua solution to the focal point. When gradient force dominates the radiation pressure of the beam then small particles could get trap near the focal point. Dielectric particles from range of nm up to μm, biological units, and polymers could be trapped and moved by this technique [32, 33].

### 3-2. Light Force, Laser Cooling

At atomic level, interaction of light and atom could be used to manipulate the atoms. The coherent interaction of the light polarizes the atom and in the presence of gradient of electromagnetic field, the atoms feel the forces. To observe this type of force, low temperature needed to avoid random thermal motion. The incoherent interaction of the light and atom can change the momentum of the atom due to direct scattering of photon from atom. The high number of photon interaction in this scenario could be used to slow down and cool the atoms. Cooling Sodium atom to 100 mK was reported by this technique in 1985 [34]. A clever approach was used to have the atoms preferentially scattering the photon from the beam opposing the direction of the motion, cooling down the atom to 240 µK [35, 36, 37]. The laser cooling was fundamental base for other important accomplishment such as Bose-Einstein condensation [38].

## 4. Electricfield Induced Manipulation

Many other approaches have been used to manipulate nanoparticles, atoms and molecules, worth mentioning electricfield induced manipulation. As an example, P. Smith et al. [39] used oscillating electricfield to positioned 370/50 nm gold nanowire in a dielectric medium between two electrodes. In more sophisticated approach, electricfield was used to transport DNA molecule in an electrolytic solution across the membrane through the nm size pore [40, 41] for detection, sizing and kinetic study.

## 5. Mechanical Manipulation

### 5-1. STM Manipulation

As mentioned in introduction, scanning probe microscopes were used for very precise manipulation of atoms and nanoparticles., Enough force between STM probe and xenon atom made Donald Eigler and Erhard Schweizer able to manipulate the atoms on nickel surface [6] at low temperature of 4 K in 1990. This work help to observe for the first time the surface electron of the copper confined by closed nano size structure made out of manipulated iron atoms by STM [42].

### 5-2-1. AFM Manipulation - Nanoparticles

First reports over manipulating of nanoparticles by AFM in controlled way was published in 1995 [43, 44]. The observation of randomly motion of the naoparticles and picking and dropping of nanoparticles during the AFM scan was reported at the same time [45]. Since then some group has used this method for fine manipulation of nano and micro particles for the experiments that high precision of distance were desired [46, 47].

At The University of Texas at Austin we have used this technique for building a precise plasmonic nano structures made of metallic and semiconductor nanoparticles. In one experiment by bringing single 35 nm Au nanosphere close to single CdSe/ZnS quantum dot, quenching of quantum dot exciton lifetime and blinking was observed as function of distance between these two nanoparticles [48]. AFM manipulation techniques had this advantage that we were able to push and move away the Au nanoparticle to and from quantum dot and monitor the effect of the distance over exciton lifetime. It is clear that other technique such as self assembly, template or DNA assembly and photolithography assembly had not the advantage of modifying the structure. In other experiment we built a nanoring structure made of four 100nm Au nanosphere and observed artificial magnetic response at visible range for the first time [49]. The magnetic signature of plasmonic structure was observed as Fano resonance due to overlapping of electric (bright mode) and magnetic (dark mode) resonances of the structure. In other work we brought the novel idea of "plasmonic protractor" to be able of reconstructing the orientation and location of a nanorod respect to a nanosphere [50]. By AFM mamnipulation technique we were able to bring 150 nm Au nanosphere within the few nm of 180/20 nm nanorod. Such a precise distance was essential to trigger the electric quadrupole response (dark mode) in the nanorod to overlap the electric dipole response (bright mode) of nanosphere.

Precision and modification of the sample is advantage of the AFM manipulation. In self assembly technique to be able to control the distance between nanoparticles, coating around nanoparticle were used as a known gap between the particles when they were come in touch in the deposition process [51,52]. After the particles were self assembled during the spin coating, the desired configuration could be found in variety of result configurations of nanoparticles. At the end particles move around and get sucked together and the gap between them are usually defined by coating of the nanoparticles. It is clear that AFM manipulation has big advantage over self assembly technique. In DNA assembly that gap is the length of the DNA between naoparticles [53].

Despite the precision advantage of AFM manipulation technique, at very short distance between naoparticles, like when you reach 5-10 nm gap, typical AFM probe do not have enough resolution to clarify that gap between the particles. So at the end AFM manipulation technique has some limit in that point. There are other challenges that all these techniques to some extend are facing them. Building 3 dimensions structure is one of that. In some DNA assembly as mention before 3 dimensions structure (like pyramid) has been reported. In AFM assembly, a not very controlled approach of pushing particles over particles has been reported [54]. We had practiced a more controlled way but more complex and time consuming of building a first layer of structure and then deposit a thin layer of buffer on the sample and build the second layer on top of that. Like coating the substrate of the assembly of 15 nm Au particles with 10 nm layer of PMMA to be able to see the position of the structure from the sticking out nanoparticles out of coating layer. Later you have to make the PMMA coating, accepting solution of nanoparticles during the spin coating to have some of the particles available on PMMA and then manipulate them to the desired area. Also we mentioned previously the report of randomly picking and dropping nanoparticles through AFM scanning. We had observed that, keeping AFM silicon probe beside metallic nanoparticles in close contact for few second bring up the chance of picking up the naoparticle. Later by rubbing the probe in different direction over some other nanoparticles we were able to drop it most of the time. This is not a control approach but an observation and option if a wanted particle was picked up accidentally.

### 5-2-2. AFM Manipulation - Atom

STM scanning was based on recording the quantum mechanics tunneling between the sharp STM probe and conductive surface. Same group that introduced STM in 1981, brought the idea of AFM in 1986 to build the topography of the surface based on the changing force between the AFM probe and the surface [55].

It took much longer for the AFM community to be able to manipulate individual atom by AFM comparing to STM [56, 57] but it was done at room temperature in comparison to low temperature STM manipulation [55]. This was lateral manipulation of Sn atom and later vertical manipulation between Sn and Si atoms was reported [58].

Scanning probe microscopy got to the point that it could even scan the chemical reaction in real time [59, 60]. Scanning probe microscopy including AFM and STM have reached to the point beyond the topography information. The boundaries cover interesting subject such as molecular interface contact, superconductivity, electron spin, plasmon field focusing, surface diffusion, bond vibration, and phase transformations [61].

## Conclusion

As it was explained, any of these popular manipulation techniques has some advantages and disadvantages and it suitable for specific application. Most of the self assembly, DNA and template assembly are suitable for making a large number of structures and when modification of the structure is not desired. STM and AFM have advantage of very precise manipulation and at level of atom manipulation. These two techniques are good for building the structures that need modification through the experiment and measurement. AFM approach is the only technique that showed ability of manipulating and moving both naoparticles and atoms. These two probe manipulating approaches are suitable for building individual complex samples. Photolithography has more precision comparing to self assembly and you can build different structure with different configuration and gap distance although the structures are not modifiable later. There is restriction on shape and materials of the nano and microstructures in lithography technique due to deposition of materials in this technique.